\documentclass[11pt]{article}
\usepackage[scale=.675,centering]{geometry}
\usepackage{amssymb,amsmath}
\usepackage{latexsym}
\usepackage[round]{natbib}
\usepackage{fleqn,latexsym}
\usepackage{smallsec,nustyle2}
\usepackage{wrapfig}
\usepackage[usenames]{color}

\renewcommand{\baselinestretch}{1.0}
\title{Deontic modality based on preference\thanks{
Contact: osherson@princeton.edu, weinstein@cis.upenn.edu}}

\author{%
Daniel Osherson \\ Princeton University \and
Scott Weinstein \\ University of Pennsylvania}

\usepackage{wasysym}
\newcounter{slic}
\newcommand{\UPS}{\addtocounter{slic}{1}}

\newcommand{\WF}[2]{\ensuremath{#1[\,#2\,]}}

\newcommand{\AGENT}{\ensuremath{\mathcal A}}
\newcommand{\MID}{\ |\ }
\newcommand{\BDL}{the basic system}

\newcommand{\BBB}{based}
\newcommand{\DBB}{$\TRI$-based}

\newcommand{\TBB}{$\TRI$-based}
\newcommand{\PDL}{the distance based system}

\newcommand{\WISS}[1]{\ensuremath{\mathsf{#1\text{-}distance}}}

\newcommand{\QED}{\ensuremath{\hfill \Box}}
\newcommand{\PROPS}{\ensuremath{\mathbb{S}}} 
\newcommand{\WORLDS}{\ensuremath{\mathbb{W}}}

\newcommand{\LANG}{\ensuremath{\mathcal{L}}}

\newcommand{\BB}{\ensuremath{\square}}
\newcommand{\BD}{\ensuremath{\lozenge}}
\newcommand{\OB}{\ensuremath{\mathbb{O}}}
\newcommand{\PE}{\ensuremath{\mathbb{P}}}
\newcommand{\CO}{\ensuremath{\mathbb{C}}}

\newcommand{\TG}{\ensuremath{\mathfrak t}} 
\newcommand{\UG}{\ensuremath{\mathfrak{u}}} 
\newcommand{\SG}{\ensuremath{\mathfrak{s}}} 
\newcommand{\WG}{\ensuremath{\mathfrak{p}}} 
\newcommand{\WGC}{\ensuremath{\mathfrak{P}}} 

\newcommand{\MO}[1]{\mbox{$\mathcal{#1}$}}

\newcommand{\TAUT}{\ensuremath{\top}}
\newcommand{\CONT}{\ensuremath{\bot}}
\newcommand{\ARGG}{\mbox{\textbf A}}
\newcommand{\AMOD}{\ensuremath{(\WORLDS, \SG, \UG, \TG)}}

\newcommand{\TRI}{\Delta}

\newcommand{\crgana}[6]{\begin{center}\fbox{\begin{minipage}{6in}   
\UPS
\textsf{Argument \arabic{slic}}: $\ARGG = #1$

\underline{Premise}: $#2$

\underline{Conclusion}: $#3$

\underline{Status}: #4

\underline{Weak status}: #5

\underline{Comment}: #6
\end{minipage}}\end{center} 
}

\newcommand{\drgana}[2]{$#1\  \models_{\TRI}\  #2$}
\newcommand{\ergana}[2]{$#1\  \models_{\;\WGC}\  #2$}
\newcommand{\grgana}[2]{$#1\  \not\models_{\;\{\WG\}}\  #2$}

\newtheorem{prop}[equation]{Proposition: }

\newtheorem{defn}[equation]{Definition: }
\newtheorem{fact}[equation]{Fact: }
\newtheorem{dispar}[equation]{}

\begin{document}
\maketitle
\begin{abstract}
\noindent
Deontic modalities are here defined in terms of the preference relation
explored in our previous work \citep{PBOR}. Some consequences
of the system are discussed.
\end{abstract}

\maketitle

\section{Introduction}\label{S:intro}

One difficulty in constructing an adequate deontic logic is the instability of
intuition about simple principles.  Let \OB\ applied to a formula represent the
obligation to render the formula true, and consider the following principle
(discussed by \citealp{Ross41}, cited in \citealp{sep-logic-deontic}).

\begin{dispar}\label{introEx1}
$\OB p \rightarrow \OB(p \vee q)$
\end{dispar}
Is \ref{introEx1} valid?
It is tempting to think so if you read the disjunction as
``at least one.'' Then the formula can be glossed as:
\begin{quote}
If you are obliged to see to it that $p$ is true then you are
obliged to see to it that at least one of $p, q$ is true.
\end{quote}

\noindent
On the other hand, \ref{introEx1} seems unpersuasive if
$p$ stands for a nice state of
affairs (you contribute to the Red Cross)
and $q$ for something bad (you pinch the cat).
In this case, $p\vee q$ does not appear to
represent a distinct obligation; at least, satisfying the disjunction with $q$ seems
not to meet any moral requirement.
[The logic advanced below will declare \ref{introEx1} to be invalid.]

Several other challenges to the development of deontic logic are discussed in
\citet[\S 4]{sep-logic-deontic}. They encourage a tolerant attitude about
alternative systems in the expectation that distinct approaches will prove
necessary in different contexts (for example, legal versus technological; see
\citealp{Meyer}).  It is in this spirit that we here offer a deontic system
built upon the \textit{preference logic} advanced in \citet{PBOR}.  The plan is
to define deontic modalities like obligation in terms of the preference
relation, then examine the consequences that follow from the underlying
preference logic. The idea of deriving obligation from preference is based on
\citet{Hansson2}, and explored more thoroughly in \citet[\S 6.2]{HanHan}.
Our system differs from earlier work because of its distinctive logic of preference.

A survey of deontic logics is available in \citet{HilMx}, including
discussion of the familiar Kripke-style semantics for obligation.
A very different approach, based on default logic,
is advanced in \citet[Ch.\ 3]{horty}.
The logic of preference is reviewed in \citet{Hansson}, and more recently in
\citet{Liu08,List}. Our system builds upon the concept of ``selecting'' a
possible world to represent a formula; this technique was introduced by
\citet{Stal} for the analysis of counterfactuals. The idea of attaching values
to possible worlds in order to analyze preference among statements (pivotal in
our system) appears in \citet{Resch} and elsewhere.  An alternative
perspective on preference is advanced in \citet{DDJLiu}.

We proceed to a summary of our approach to preference logic, then return to
deontic modality. The appendix sketches the proofs of the
propositions that follow.

\section{A preference logic}\label{S:plog}

We consider only the most elementary system formulated in \citet{PBOR}, leaving
generalizations to one side. Also, the non-modal basis of the logic will be
limited to the sentential calculus; extension to quantifiers is discussed in
\citet{pbbook}.

\subsection{Language}\label{SS:lang}

The language \LANG\ of our preference logic is built from a nonempty set
\PROPS\ of sentential variables, the unary connective $\neg$, the binary
connectives $\wedge$ and $\succeq$, along with the two parentheses.  Formulas
are defined inductively via:

\def\SPACEA{\hspace{1.65mm}}
\begin{center}
$p\in\PROPS \SPACEA \mid \SPACEA \neg \varphi \SPACEA \mid
\SPACEA (\varphi \wedge \psi) \SPACEA \mid \SPACEA
(\varphi\succeq \psi)$ 
\end{center}

\noindent
We rely on the following abbreviations.
\def\ABB{\text{for}}
\begin{eqnarray*}
(\varphi\vee\psi) & \ABB & \neg(\neg\varphi \wedge\neg\psi)\\
(\varphi\rightarrow\psi) & \ABB & (\neg\varphi \vee\psi)\\
(\varphi\leftrightarrow\psi) & \ABB & ((\varphi
\rightarrow\psi)\wedge (\psi \rightarrow\varphi))\\
(\varphi\succ \psi) & \ABB & (\varphi\succeq\psi)\wedge
\neg(\psi\succeq\varphi)\\
(\varphi\approx\psi) & \ABB & (\varphi\succeq\psi)\wedge
(\psi\succeq\varphi)\\
(\varphi\preceq\psi) & \ABB & (\psi\succeq\varphi)\\
(\varphi\prec\psi) & \ABB & (\psi\succ\varphi)\\
\TAUT & \ABB & (p \rightarrow p)\\
\CONT & \ABB & \neg\TAUT\\
\end{eqnarray*}

To perceive the intended meaning of $\varphi \succ \psi$, fix
an agent \AGENT\ whose reasoning is at issue. Then $\varphi \succ \psi$
is true if and only if

\def\MPS{4.25in}
\begin{center}
\begin{minipage}{\MPS}
\AGENT\ envisions a situation in which $\varphi$ is true and that otherwise
differs little from her actual situation (if $\varphi$ is already true then
\AGENT's actual situation may well be the one she envisions). Likewise, \AGENT\
envisions a second situation that is like her actual situation except that
$\psi$ is true. Finally, the utility of the first imagined situation exceeds
that of the second.
\end{minipage}
\end{center}

\noindent In our formal semantics, the appeal to utility is underwritten by a
function that maps possible worlds into real numbers.  Several generalizations
involving just an ordering of worlds are explored in \citet{PBOR}.  Also, the
latter paper represents utility as fractionated into separate scales thought of
as reasons for preference. For example, if \AGENT\ is contemplating a
political career, different considerations might compete in her mind, leading
to a final evaluation of the options \textit{proceed} or \textit{desist}. To
keep the present system as simple as possible, we here dispense with the issue
of utility aggregation (explored in \citealp{PBOR,pbbook}), letting $\succeq$
embody relative value ``all things considered.'' 

The appeal to alternative situations is based on a \textit{selection function}
of the kind envisioned by \citet{Stal}, namely, as mapping a
world-plus-proposition into another world that satisfies the proposition.
Intuitively, the chosen world is ``similar'' to the original world; it might
even \textit{be} the original world if the proposition is true in it. To
enforce similarity, we will ultimately impose a powerful condition on selection
functions.

Note that \LANG\ allows embedded occurrences of $\succ$ as in: $p \succ (q
\succ r)$. If $p$ is loneliness, $q$ is amassing great wealth, and $r$ is
finding true love then the formula would be true of \AGENT\ if she would rather
be lonely than prefer riches to love (a real romantic).

\subsection{Models}\label{SS:sem}

Models for \LANG\ are built from a nonempty set \WORLDS\ that embodies the
imaginative possibilities (``worlds'') available to an agent in the course of
practical deliberation. Subsets of \WORLDS\ are called \textit{propositions}.
Selection functions are defined as follows.

\begin{defn}\label{selfnc}
A \textit{selection function \SG\ over \WORLDS} is a mapping
from $\WORLDS\times\{A\subseteq\WORLDS\mid A\neq \emptyset\}$
to \WORLDS\ such that for all $w\in\WORLDS$ and $\emptyset\neq
A\subseteq\WORLDS$, $\SG(w,A)\in A$.
\end{defn}

\noindent Thus, $\SG(w,A)$ is a choice of world to represent $A$, where the
choice depends on $w$.  Next, each world is evaluated according to
the model's utility function.

\begin{defn}\label{util}
A \textit{utility function over \WORLDS} is a mapping
from \WORLDS\ to the real numbers.
\end{defn}

\noindent Recall that \PROPS\ is the set of sentential variables
in \LANG. The last component of a model is an assignment of proposition to
each variable in \PROPS.

\begin{defn}\label{tru1}
A \textit{truth-assignment over \WORLDS} is a
mapping from $\PROPS$ to the power set of $\WORLDS$.
\end{defn}
\noindent
Models are defined as follows.

\begin{defn}\label{mod1}
A \textit{model} is a quadruple \AMOD\ where
\begin{enumerate}
\item \WORLDS\ is a nonempty set of worlds;
\item \SG\ is a selection function over \WORLDS;
\item \UG\ is a utility function over \WORLDS;
\item \TG\ is a truth-assignment over \WORLDS;
\end{enumerate}
\end{defn}

\subsection{Semantics}\label{SS:tru}

It remains to specify the proposition (set of worlds)
expressed by a formula $\varphi$ in a model \MO{M}. This
proposition is denoted \WF{\varphi}{\MO{M}}, and defined
inductively as follows.

\begin{defn}\label{defSat2}
Let $\varphi\in\LANG$ and model $\MO{M} = \AMOD$ be given.
\begin{enumerate}
\item If $\varphi \in \PROPS$ then $\WF{\varphi}{\MO{M}} = \TG(\varphi)$.
\item If $\varphi$ is the negation $\neg\theta$ then
$\WF{\varphi}{\MO{M}} = \WORLDS \setminus \WF{\theta}{\MO{M}}$.
\item If $\varphi$ is the conjunction $(\theta\wedge\psi)$ then
$\WF{\varphi}{\MO{M}} = \WF{\theta}{\MO{M}} \cap \WF{\psi}{\MO{M}}$.
\item\label{defSat2a}
If $\varphi$ has the form $(\theta\succeq\psi)$ then
$\WF{\varphi}{\MO{M}} = \emptyset$ if either
$\WF{\theta}{\MO{M}} = \emptyset$ or $\WF{\psi}{\MO{M}} = \emptyset$. Otherwise:
\[\WF{\varphi}{\MO{M}} = \{w\in\WORLDS\mid \UG(\SG(w,\WF{\theta}{\MO{M}})) \ge
\UG(\SG(w,\WF{\psi}{\MO{M}}))\}.\]
\end{enumerate}
\end{defn}

\noindent Observe that \WF{(\theta\succeq\psi)}{\MO{M}} is
defined to be empty if there is no world that satisfies
$\theta$ or none that satisfies $\psi$. Thus, we read
$(\theta\succeq\psi)$ with existential import (``the
$\theta$-world is weakly better than the $\psi$-world,''
where the definite description is Russellian). In the nontrivial
case, let $A\neq\emptyset$ be the proposition expressed by
$\theta$ in \MO{M}, and $B\neq\emptyset$ the one expressed by
$\psi$. Then (intuitively) world $w$ satisfies
$(\theta\succeq\psi)$ in \MO{M} iff the world selected from $A$
as closest to $w$ has utility no less than that of the
world selected from $B$ as closest to $w$.
%

Let $\varphi,\psi\in\LANG$ be given.
Then (in the standard way), we say that
$\varphi$ \textit{implies} $\psi$ ($\varphi\models\psi$)
just in case for every model
$\MO{M} = \AMOD$, $\WF{\varphi}{\MO{M}} \subseteq \WF{\psi}{\MO{M}}$.
Also, $\varphi$ is \textit{valid} ($\models \varphi$) just in case
for every model
$\MO{M} = \AMOD$, $\WF{\varphi}{\MO{M}} = \WORLDS$.
To illustrate, it is easy to show that for all $\varphi,\psi,\theta\in \LANG$:
\begin{itemize}
\item[] $\models \varphi \succeq \varphi$
\item[] $(\varphi \succeq \psi)\wedge (\psi \succeq \theta)\ \models\  \varphi \succeq \theta$
\item[] $\models \neg(\CONT\succeq \varphi)$ \quad\ and \quad\ $\models \neg(\varphi\succeq \CONT)$
\end{itemize}

\noindent The foregoing definitions relativize in the obvious way to subclasses
of models. For example, $\varphi$ implies $\psi$ in a given class $C$ of models
just in case for every model $\MO{M} \in C$, $\WF{\varphi}{\MO{M}}
\subseteq \WF{\psi}{\MO{M}}$.  In this case, we write $\varphi\models_C\psi$.

\subsection{Alethic modality}\label{SS:am}

Our logic allows expression of the global modality (see \citealt[\S 2.1]{Blackburn}).
For $\varphi\in\LANG$, let:
\begin{dispar}\label{boxdef}
\[
\BB\varphi \overset{\normalfont{\text{def}}}{=}
\neg(\neg\varphi \succeq\neg\varphi) \quad\ \text{and} \quad\ 
\BD\varphi \overset{\normalfont{\text{def}}}{=}
(\varphi \succeq\varphi).
\]
\end{dispar}
\noindent
Then unwinding clause \ref{defSat2}\ref{defSat2a} of our semantic definition
yields:
\begin{prop}\label{global}
For all $\varphi\in\LANG$ and models $\MO{M} = \AMOD$:
\begin{enumerate}
\item\label{globala}
$\WF{\BB\varphi}{\MO{M}}\neq\emptyset$ iff $\WF{\BB\varphi}{\MO{M}} =
\WORLDS$ iff $\WF{\varphi}{\MO{M}} = \WORLDS$.
\item\label{globalb}
$\WF{\BD\varphi}{\MO{M}}\neq\emptyset$ iff $\WF{\BD\varphi}{\MO{M}} =
\WORLDS$ iff $\WF{\varphi}{\MO{M}} \neq\emptyset$.
\end{enumerate}
\end{prop}
From the proposition, we obtain that the axioms of S5 are valid for \BB\ and \BD.
Here are some other results involving modality.
For all $\varphi,\psi,\theta\in \LANG$:
\begin{itemize}
\item[] $\models\BB\varphi \leftrightarrow \neg\BD\neg\varphi
\quad\quad \models\BD\varphi \leftrightarrow \neg\BB\neg\varphi$
\item[] $\BD\varphi \wedge \BD\psi\ \models\ (\varphi \succeq \psi)\vee (\psi \succeq \varphi)$
\item[] $\models(\BD\varphi \wedge \BD\psi)\leftrightarrow
(\neg(\varphi \succeq \psi)\leftrightarrow (\psi \succ \varphi))$
\item[] $\models \BD\varphi \rightarrow (\varphi \approx \psi)$ if
$\models(\varphi\leftrightarrow \psi)$.
\end{itemize}

\section{A deontic system}\label{Sec:DS}

We now offer definitions of deontic modalities in terms of $\succeq$. The
definitions will be successful if they imply a rich set of principles about
obligation while avoiding dubious claims.

\subsection{The primacy of conditional obligation}\label{Sec:CO}

\def\SONE{Life expectancy rises.}
\def\STWO{Social Security revenues are enhanced.}
Suppose that $p, q$ stand for the sentences below.

\begin{dispar}\label{hunger1}
\begin{description}
\item[$p$\,:\ ] \SONE
\item[$q$\,:\ ] \STWO
\end{description}
\end{dispar}

\noindent Then a situation satisfying $p$ ought to satisfy $q$ as well
(according to progressives). Let us denote this judgment by $\CO(p,q)$, where
\CO\ is the dyadic modality of conditional obligation. It might be thought that
monadic obligation (denoted by \OB) is more fundamental than \CO, allowing the
latter to be defined in one of the ways shown here:

\begin{dispar}\label{baddef}
\[\text{(a)}\hspace*{.2in} \CO(p,q) \overset{\normalfont{\text{def}}}{=}
\OB(p\rightarrow q) \] \[\text{(b)}\hspace*{.2in} \CO(p,q)
\overset{\normalfont{\text{def}}}{=} p\rightarrow \OB q \]
\end{dispar}

\noindent But neither definition is satisfactory. Since $p \rightarrow q$ is
equivalent to $\neg q \rightarrow \neg p$, the first suggests that $\CO(p,q)$
if and only if $\CO(\neg q,\neg p)$. Letting $p, q$ be given by \ref{hunger1},
we see that, to the contrary, $\CO(p,q)$ can be true and $\CO(\neg q,\neg p)$
false.  For in a situation where Social Security revenues are not enhanced
we are not obliged to lower life expectancy. The second definition in \ref{baddef}
entails that $C(p,q)\vee C(\neg p, q)$ is a logical truth, which is odd.
(Neither attending baseball games nor failing to requires that I root for the
Red Sox.)

More discussion of the obstacles to representing conditional obligation by a
composite of \OB\ and the material conditional is available in \citet[\S
4.5]{sep-logic-deontic}. We here follow the consensus opinion (first
articulated in \citealp{Chis}) that conditional obligation is the fundamental
deontic concept, monadic obligation being derivative to it.

\subsection{Deontic definitions}\label{Sec:dd}

So let us begin with conditional obligation.

\begin{defn}\label{codef} For all $\varphi,\psi\in\LANG$, \[ \CO(\varphi,\psi)
\overset{\normalfont{\text{def}}}{=} (\varphi \wedge \psi) \succ (\varphi
\wedge \neg\psi) \] \end{defn}
Continuing with the example \ref{hunger1}, Definition \ref{codef} glosses
$\CO(p,q)$ as:

\begin{center} \begin{minipage}{\MPS}
Life expectancy rises and Social Security revenues are enhanced
\textit{is better than} 
life expectancy rises and Social Security revenues are not enhanced.
\end{minipage} \end{center}

\noindent The obligation is conditional in the sense that neither $p$ nor $q$
is rendered obligatory by $\CO(p,q)$.
We will not attempt to survey all the
alternatives to \ref{codef} that come to mind but are unsuitable.
One illustration suffices. The definition
\begin{quote} \[ \CO(\varphi,\psi)
\overset{\normalfont{\text{def}}}{=} \neg\varphi \succ (\varphi
\wedge \neg\psi) \] \end{quote}

\noindent seems promising at first. ($\varphi$ carries committment to $\psi$ if
it's better that $\varphi$ be false than be true without $\psi$.) But consider:

\begin{quote}
\begin{description}
\item[$p$\,:\ ] You save someone from drowning.
\item[$q$\,:\ ] You exhibit modesty when subsequently interviewed.
\end{description}
\end{quote}

\noindent For these sentences, $\CO(p,q)$ is plausible but $\neg p \succ
(p\wedge \neg q)$ is not since it's better to save someone and brag about it
than to let the victim drown.

Definition \ref{codef}, as well as its competitors, is open to concerns about
relevance.  If $p$ is tying your shoelaces, and $q$ is saving a rainforest then
$p\wedge q$ is better than $p\wedge \neg q$. Yet it seems strange to believe
that $p$ carries the obligation $q$. It is not the mission of our logic,
however, to keep track of all the circumstances giving rise to obligations;
there may well be a back story that connects your shoelaces to the fate of a
rain forest.  Definition \ref{codef} should rather be assessed in terms of the
plausibility of its consequences on the assumption that $p$ carries the
obligation $q$. Moreover, in a given application of deontic logic, all the concepts
in play might be relevant to each other.

Following \citet[p.\ 171]{Hilp}, the
two monadic modalities (obligation and permission) may be derived from $\CO$.

\begin{defn}\label{obdef} For all $\psi\in\LANG$, \[ \OB\psi
\overset{\normalfont{\text{def}}}{=} \CO(\TAUT,\psi) \hspace*{.5in} \PE\psi
\overset{\normalfont{\text{def}}}{=} \neg\OB\neg\psi \] \end{defn} Thus, $\psi$
is obligatory if it is required by the logically true proposition. $\psi$ is
permissible if its negation is not obligatory.  Unravelling the definitions
yields: \begin{prop}\label{unrav1} For all $\psi\in\LANG$, \begin{enumerate}
\item\label{unrav1a} $\models(\OB \psi) \leftrightarrow (\psi\succ\neg\psi)$
\item\label{unrav1b} $\models(\PE \psi) \leftrightarrow (\psi\succeq\neg\psi)$
\end{enumerate} \end{prop} That is, $\psi$ is obligatory if its truth makes for
a better world than does its falsity; $\psi$ is permitted provided its truth
doesn't make things worse.
The semantics of $\succeq$ presented in Section \ref{SS:tru}, along with
Definitions \ref{codef} and \ref{obdef}, will be called \textit{\BDL}.
Easy consequences of \BDL\ include:

\def\XSP{\hspace*{.15in}}
\begin{prop}\label{ST} For all $\theta,\psi\in\LANG$,
\begin{enumerate}
\item\label{ST6} $\BD(\theta) \wedge \BD(\neg \theta) \models \PE(\theta) \vee \PE(\neg\theta)$
\item\label{ST4} $\OB\theta\ \models\ \PE\theta$ 
\item\label{ST1} $\models\neg((\OB \theta) \wedge \OB(\neg\theta))$ 
\item\label{ST3} $\models\neg \OB\TAUT$ \XSP $\models\neg\OB\CONT$ \XSP $\models\neg\PE\CONT$ \XSP $\models\neg\PE\TAUT$
\item\label{ST5} $\CO(\theta,\psi)\ \models\ \BD(\theta\wedge\psi)\wedge \BD(\theta\wedge\neg\psi)$.
\end{enumerate}
\end{prop}

\noindent
Glosses for \ref{ST} are straightforward; for example,
\ref{ST}\ref{ST5} asserts that conditional obligations involve contingent propositions.
The same is true for monadic obligation, as seen in
\ref{ST}\ref{ST3} --- which already distinguishes \BDL\ from
deontic logic based on Kripke models.
The principles in \ref{ST} can be questioned but they are not outlandish.
Nonetheless, the harvest of
deontic principles in \ref{ST} is rather meager.  Richer results will emerge in Section
\ref{SS:prox} when we relativize validity to a narrower class of models.

\subsection{Axiomatization}\label{Sec:ax}

The weakness of \BDL\ is reflected in the simplicity of the formulas needed to
axiomatize it.  Axioms include all instances of any standard schematic
axiomatization of S5 [using the modality defined in \ref{boxdef}], together
with all instances of the following additional schemata.

\begin{dispar}\label{axioms}
\begin{enumerate}
\item $((\varphi \succeq \psi)\wedge (\psi \succeq \theta))\rightarrow (\varphi \succeq \theta)$\label{axtrans}
\item $(\BD\varphi \wedge \BD\psi)\leftrightarrow ((\varphi \succeq \psi)\vee (\psi \succeq \varphi))$\label{axconn}
\item $\BB(\varphi \leftrightarrow \psi)\rightarrow (((\varphi \succeq \theta)\leftrightarrow (\psi \succeq \theta))\wedge((\theta \succeq \varphi)\leftrightarrow ( \theta\succeq \psi)))$\label{axext}
\end{enumerate}
\end{dispar}

\noindent The axioms express little more than the preordering imposed by
$\succeq$, a substitution property, and the S5 apparatus.  The set of theorems
is obtained from the closure of the axioms under \textit{modus ponens} and
necessitation.  Of course, the deontic content of a theorem must be extracted
via Definitions \ref{codef} and \ref{obdef}.

It follows from results demonstrated in \citet[\S 3]{pbbook} that the set of
theorems is exactly the class of valid formulas of \BDL. Proofs of the
decidability of this class, along with other metatheorems, are available in
\citet[\S 6]{PBOR} and \citet[\S 4]{pbbook}.

\subsection{Chisolm's paradox}\label{SS:chisP}

We pause to consider \textit{Chisolm's paradox}
(reviewed in \citealp[\S 4.5]{sep-logic-deontic})
which turns on the following statements.

\begin{dispar}\label{chispar}
\begin{enumerate}
\item It ought to be that Jones \underline{g}oes to the assistance of his neighbors.
\item It ought to be that if Jones \underline{g}oes then he \underline{t}ells them he is coming.
\item\label{chispar3} If Jones doesn't \underline{g}o then he ought not \underline{t}ell them he is coming.
\item Jones doesn't \underline{g}o.
\end{enumerate}
\end{dispar}

\noindent
A straightforward formalization of \ref{chispar} in terms of \OB\ and the
material conditional --- for example, statement \ref{chispar}\ref{chispar3} as
$\neg g \rightarrow \OB\neg t$ --- leads to contradiction within the usual
Kripke style semantics for deontic logic. (For Kripke semantics, see
\citealp[Ch.\ 6]{Chellas}).  This is unwelcome inasmuch as the four statements seem
perfectly compatible.  In the present system, however, \ref{chispar} goes over
to:

\begin{dispar}\label{chisparUs}
\begin{enumerate}
\item $\OB g$
\item $\CO(g,t)$
\item $\CO(\neg g,\neg t)$
\item $\neg g$
\end{enumerate}
\end{dispar}
The formulas in \ref{chisparUs} are satisfiable not only in \BDL\
but also in the class of models that we now introduce.

\section{Models based on symmetric difference}\label{SS:pdl}

A class of models will now be introduced that gives sharper meaning to the idea
that an envisioned world is ``close'' to the actual one.

\subsection{$\TRI$ models}\label{SS:prox}

Recall that \PROPS\ is the set of propositional variables that underlie our
modal language \LANG. We henceforth identify the class $\WORLDS$ of worlds in a
model with the power set of \PROPS.  Also, the map $\TG$ from variables to
propositions will always be given by:
\begin{quote}
$\TG(p) = \{w\in\WORLDS\MID p \in w\}$, for every $p\in\PROPS$.
\end{quote}
\noindent That is, $p$ is mapped to the set of worlds that include $p$ as a member.

The \textit{symmetric difference} between sets $X,Y$ is standardly denoted
$X\TRI Y$ and defined to be $(X\setminus Y) \cup (Y\setminus X)$.  Thus,
the disagreement between two worlds $w_0,w_1$ may be represented by $w_0\TRI
w_1$.  We call a selection function $\SG$ \textit{\TBB} if for all
$w_0\in\WORLDS$ and propositions $A$ there is no $w_1\in A$ with $w_0\TRI
w_1\ \subset\ w_0\TRI \SG(w_0,A)$. (Here, $\subset$ is used in the
``proper'' sense.)
 
To illustrate \TBB\ selection, let $A = \{w\in\WORLDS\MID p\in w,\; q\not\in
w\}$.  Suppose that world $w_0$ includes both $p$ and $q$.  Then \TBB\
$\SG(w_0,A)$ returns the world just like $w_0$ except that $q$ is excluded. In
particular, $r\in\SG(w_0,A)$ iff $r\in w_0$, and likewise for all other variables
distinct from $p, q$. If $w_1$ excludes both $p$ and $q$ then $\SG(w_1,A)$ returns
the world just like $w_1$ except that $p$ is present. If $w_2$ includes $p$ but
not $q$ then $w_2\in A$ and $\SG(w_2,A) = w_2$. (In this sense, \TBB\ selection
is ``reflexive.'') If $B$ is the set of worlds that includes either $p$ or $q$
or both, and $w_3$ excludes both $p$ and $q$ then \SG\ must make a choice.
Either $\SG(w_3,B)$ is just like $w_3$ except for including $p$, or
$\SG(w_3,B)$ is just like $w_3$ except for including $q$.  Note that
$\SG(w_3,B)$ cannot return a world in which both $p$ and $q$ are present
because $B$ includes closer alternatives to $w_3$, for example, the world just
like $w_3$ except for including $p$. 

Summarizing:

\begin{defn}\label{proxM}
A model \AMOD\ is called $\TRI$ just in case
\begin{enumerate}
\item\label{proxHa} \WORLDS\ is the power set of \PROPS;
\item\label{proxHc} for all $p\in\PROPS$,
$\TG(p) = \{w\in\WORLDS\MID p \in w\}$;
\item\label{proxHb} \SG\ is \TBB.
\end{enumerate}
\end{defn}
Observe that $\TRI$ models
place no restriction on the utility function \UG; it remains
an arbitrary mapping of \WORLDS\ to the reals.

\subsection{Validity in $\TRI$}\label{SS:tant}

All the facts about \BDL\ that are recorded in Proposition
\ref{ST} carry over when relativized to $\TRI$.  We also have
the following facts (not derivable in the basic system).

\begin{prop}\label{DBProp1}
\begin{enumerate}
\item\label{DBProp1f}
\drgana{\OB p \wedge (\neg p \succ\neg q)}{\OB q}
\item\label{DBProp1e}
\drgana{\OB q \wedge (p \succ q) }{\OB p}
\item\label{DBProp1d}
\drgana{\CO(p,q)\wedge p}{\OB q}
\item\label{DBProp1a}
\drgana{\CO (p \wedge r, q) \wedge \CO (p \wedge \neg r, q)}{\CO(p,q)}
\item\label{DBProp1b}
\drgana{\CO( p, q)}{\CO (p \wedge r, q) \vee \CO (p \wedge \neg r, q)}
\item\label{DBProp1c}
\drgana{\OB(p \wedge q) \wedge (\OB(p \wedge \neg q)}{\OB p}
\end{enumerate}
\end{prop}

\noindent
Fact \ref{DBProp1}\ref{DBProp1f} is close to the ``contranegativity'' principle
discussed in \citet{Hansson2}. Hansson's principle is plausible: If $p$ is
obligatory while missing out on $q$ is worse than missing out on $p$ then $q$
should also be obligatory.  The positive version \ref{DBProp1}\ref{DBProp1e}
seems equally reasonable.  Fact \ref{DBProp1}\ref{DBProp1d} is an expected
detachment principle for conditional obligation. The next two facts can be
interpreted as factorization properties of conditional obligation. Likewise,
\ref{DBProp1}\ref{DBProp1c} expresses factorization for monadic obligation.

But notice that we are missing the analog of \ref{DBProp1}\ref{DBProp1b} for
monadic obligation --- namely, the inference from $\OB p$ to $\OB(p \wedge q)
\vee (\OB(p \wedge \neg q)$ --- which would complement
\ref{DBProp1}\ref{DBProp1c}.  Indeed, the foregoing inference is invalid in
$\TRI$.  Likewise, $\TRI$ fails to validate other attractive
principles, such as the inference from $\CO(p,q)\wedge\CO(r,q)$ to $\CO(p\vee
r,q)$. In response, we now introduce a weaker criterion of validity which
performs better at classifying deontic principles.

\section{$\TRI$-models that incorporate weight}

\subsection{Weighting functions}

The distance between worlds $w_0,w_1$ has so far been measured qualitatively in
terms of symmetric difference. A more nuanced reckoning would allow some
variables to affect distance more than others.  For this purpose, we define a
\textit{weighting function} to be any map of \PROPS\ (the set of sentential
variables) to the positive real numbers. In the context of such a map, the distance
between two worlds may be defined as follows.

\begin{defn}\label{weights1}
Let weighting function \WG, and worlds $w_0,w_1\subseteq\PROPS$ be given.
Then:
\[
\WISS{\WG}(w_0,w_1)\quad \overset{\normalfont{\text{def}}}{=}\quad
\sum_{v \in w_0\TRI w_1} \WG(v).
\]
\end{defn}
\noindent
Thus, \WISS{\WG} is given by the sum of the weights of each variable
that has different membership status in the two worlds at issue. (The \WG\
stands for \textit{poids}, the use of $w$ inviting confusion with worlds.) 

\subsection{Selection functions sensitive to weight}

Let $\WG$ be a weighting function.
A selection function 
$\SG$ is said to be \textit{\WG\text{-}\BBB} if for all
$w\in\WORLDS$ and nonempty $A\subseteq\WORLDS$, $\SG(w,A)$ is chosen from among
the \WG-nearest worlds to $w$ in $A$. In other words, for \SG\ to be \WG-\BBB, there
must be no $w'\in A$ with $\WISS{\WG}(w,w') < \WISS{\WG}(w,\SG(w,A))$.
Intuitively, the greater the weight that \WG\ attaches to a variable $q$,
the more reluctant is a \WG-\BBB\ selection function to return a world in which
$q$ has changed its truth-value.
Since weighting functions return positive numbers, it is easy to verify:
\begin{fact}\label{smoothFact}
Every \WG-\BBB\ selection function is \TBB.
\end{fact}

\subsection{\WGC-validities}

Because of \ref{smoothFact}, weighting functions can be used to define subclasses
of $\TRI$ models.

\begin{defn}\label{cmwf}
Let \WGC\ be a class of weighting functions. For $\varphi,\psi\in\LANG$,
we write $\varphi\models_{\;\WGC}\psi$ just in case for every
$\WG\in\WGC$, every \WG-\BBB\ selection function \SG, and
every $\TRI$ model \MO{M} with \SG\ as selection function,
$\WF{\varphi}{\MO{M}}\subseteq \WF{\psi}{\MO{M}}$.
\end{defn}
\noindent
As a special case,
$\emptyset\models_{\WGC}\psi$ just in case for every
$\WG\in\WGC$, every \WG-\BBB\ selection function \SG, and
every \TBB\ model $\MO{M}$ with \SG\ as selection function,
$\WF{\psi}{\MO{M}} = \WORLDS$.

It is easy to see that the validities appearing in Propositions \ref{ST} and
\ref{DBProp1} are also valid in the sense of Definition \ref{cmwf}, for any
class \WGC\ of weighting functions. If \WGC\ is chosen with care, we
obtain further principles, as in the following. 

\begin{prop}\label{wv}
Let \WGC\ be the class of weighting functions that assign greater weight
to $q$ than to $p$ and to $r$. Then:
\begin{enumerate}
\item\label{wva} \ergana{\CO(r, p)}{\CO (r, p\wedge q) \vee \CO (r, p\wedge \neg q)}
\item\label{wvb} \ergana{\OB p}{\OB(p \wedge q) \vee (\OB(p \wedge \neg q)}
\item\label{wvc} \ergana{\PE p}{\PE(p \wedge q) \vee \PE(p \wedge \neg q)}
\item\label{wvd} \ergana{\PE(p \wedge q) \wedge (\PE(p \wedge \neg q)}{\PE p}
\item\label{wve} \ergana{\CO(q,p)\wedge \CO(r,p)}{\CO(q\vee r, p)}
\item\label{wvf} \ergana{\CO(p\vee q,r)}{\CO(p,r)\vee \CO(q,r)}
\item\label{wvg} \ergana{\OB(p\wedge q)\wedge p}{\OB q}
\item\label{wvh} \ergana{\OB(p\vee q)\wedge \neg q}{\OB p}
\end{enumerate}
\end{prop}

\noindent
If the variables in these principles are permuted, the class \WGC\ of
weighting functions must be adjusted accordingly. The proposition only
informs us that each of the principles is reliable in a broad and simple
class of weighting functions, reflecting the relative importance of
conserving the polarity of a certain variable when exploring possible
worlds. It is hoped that this goes some way towards explaining the
deontic appeal of the principles in the proposition.

Indeed, \ref{wv}\ref{wva}-\ref{wvd} express plausible fractionation properties
of conditional obligation, monadic obligation, and permission; they complement
the list provided by Proposition \ref{DBProp1}. Item \ref{wv}\ref{wve} commits
an agent to leaving a tip if she goes to either of two fine restaurants on the
assumption that she should leave a tip if she goes to the first and leave a tip
if she goes to the second. This seems reasonable. Item \ref{wv}\ref{wvf} is a
partial converse. Suppose that the agent is comitted to paying U.S.\ taxes on
the assumption that she lives in one of Chicago or Moscow. Then either she is
commited to paying U.S.\ taxes on the assumption of living in Chicago or she is so
committed on the assumption of living in Moscow.  Concerning \ref{wv}\ref{wvg},
if Johnny is obliged to feed the cat and straighten his room but the cat is
already fed, then Johnny is left with the obligation to straighten his room.
To illustrate \ref{wv}\ref{wvh}, if the government must either build a tunnel
or build a bridge and a bridge is not built, then the government must build a
tunnel.

\subsection{Invalidity}

The facts presented above suggest that our logic validates many deontically
acceptable principles, or at least validates them relative to a large and
coherent class of weighting functions. But a satisfactory deontic logic
is also required to mark unacceptable principles as invalid.
The natural choice for such a mark is invalidity with respect to every
weighting function.
We then have the following.

\begin{prop}\label{sk} For every weighting function \WG: \begin{enumerate}
\item\label{ska}\grgana{\CO(p\vee q,r)}{\CO(p,r)\wedge \CO(q,r)} 
\item\label{skb}\grgana{\OB(p\wedge q)}{\OB q} 
\item\label{skc}\grgana{\OB(p\vee q)\wedge \OB(\neg q)}{\OB p} 
\item\label{skd}\grgana{\CO(p, q)}{\CO(p\wedge r,q)}
\item\label{ske}\grgana{\CO(p,q)\wedge \OB p}{\OB q} 
\item\label{skz}\grgana{\OB(p\rightarrow q)}{\OB(p) \rightarrow\OB(q)}
\item\label{skf}\grgana{\CO(p, q)\wedge \CO(q, r)}{\CO(p, r)}
\item\label{skg}\grgana{\OB p\wedge\OB q}{\OB (p\wedge q)}
\item\label{ski}\grgana{\CO(p,q)\wedge \CO(p,r)}{\CO(p, q \wedge r)}
\item\label{skj}\grgana{\OB\OB p}{\OB p}
\item\label{skk}\grgana{\CO(p,q)\wedge \CO(r,q)}{\CO(p\wedge r, q)} 
\item\label{skm}\grgana{\emptyset}{\OB(\OB p \rightarrow p)}
\item\label{skh}\grgana{\OB p}{\OB (p\vee q)}
\end{enumerate} \end{prop}

\noindent Let us try to jutify the items in \ref{sk}.
The same illustration used before shows that strengthening \ref{wv}\ref{wvf} to
\ref{sk}\ref{ska} fails.  Being comitted to paying U.S.\ taxes on the
assumption that you live in one of Chicago or Moscow leaves open the
possibility that the commitment arises from living in Chicago, and this does
not entail a commitment to paying U.S.\ taxes if one lives in Moscow.
Similarly, \ref{wv}\ref{wvg} cannot be strengthened to \ref{sk}\ref{skb}. A
medical team may have the obligation to apply anesthetic and make an incision.
This does not engender the unqualified obligation to make an incision for,
despite their obligations, they might not have applied anesthetic.
The same consideration underlies the inability to rewrite \ref{wv}\ref{wvh} as
\ref{sk}\ref{skc}.  Suppose you're obliged to throw either Tom or Harry out of
the life boat. Because of his saintly character, you're obliged not to do this
to Harry. But it does not follow that you're obliged to throw Tom into the sea.
It depends on whether you fulfilled your obligation to Harry. If Harry has
already been thrown overboard (despite your obligation), then you're not
required to make Tom join him. 

Item \ref{sk}\ref{skd} is undermined by setting $p$ = I'm invited to a party
honoring a close friend, $q$ = I go to the party, and $r$ = the police advise
me not to go. For \ref{sk}\ref{ske}, consider $p$ = I go to church, and $q$ = I
shout hallelujah.  Although it may be true that going to my particular church
obliges me to shout hallelujah, and that I am obliged to go to church, it
doesn't follow that I'm obliged to shout hallelujah, for I might not have
fulfilled my obligation to go to church (and shouting hallelujah outside of
church annoys people).  The contrasting case with $p$ rather than $\OB p$ is
listed among the $\TRI$-validities as \ref{DBProp1}\ref{DBProp1d}, above.
Item \ref{sk}\ref{skz} is open to the same kind of counter example as
\ref{sk}\ref{ske}; we rehearse it in the Discussion section in light of
the centrality of the principle to standard deontic logic.

A similar consideration undermines \ref{sk}\ref{skf}, a putative transitivity
principle for conditional obligation. Suppose that an invasion by the enemy
obliges us to stand and fight, and that the latter obliges us to blow
up the bridges. It doesn't follow that an invasion obliges us to blow up
the bridges.  It depends on whether we stand and fight or capitulate, that
is, whether we live up to our first conditional obligation. If not, we
shouldn't blow up the bridges, violating transitivity.

Item \ref{sk}\ref{skg} raises the issue of contradictory obligations. Perhaps a
person could have the obligation to spend Saturday morning at the office (due
to unfinished work that people are waiting for), and also the obligation to
spend Saturday morning with her children.  We might nonetheless resist the idea
that the impossible conjunction of the two duties is obligatory. Such is the
view of the logic developed here, which restricts obligation to contingent
propositions.  [Cf.\ \ref{ST}\ref{ST3}].  Similar remarks apply to the related
principle \ref{sk}\ref{ski}. For deeper discussion of moral conflict, see
\citet[Ch.\ 4]{horty}.

The modal collapse envisioned in \ref{sk}\ref{skj} does not seem deontically
viable. Let $p$ = No one consumes more than $500$ calories per day.  It would
be good if that were good since we would save on groceries.  But it would not
be good that $p$, since that would leave everyone malnourished.  Regarding
\ref{sk}\ref{skk}, let $p$ be that you promise marriage to Joan, $r$ be that
you promise marriage to Jane, and $q$ that you go to City Hall to swear
monogamy. Each promise separately requires the oath of monogamy but their
conjunction seems not to.  Next, the formula $\OB(\OB p \rightarrow p)$
appearing in \ref{sk}\ref{skm} is initially attractive; it says that
it is good that if $p$ is good then $p$ is true. But let $p$ be that
global temperature is stable, and consider the equivalent formulation:
$\OB(\neg p \rightarrow \PE \neg p)$ --- which says (dubiously)
it is good that if temperatures are unstable then this is permissible.
Moreover, there might be conflicting good things as in $p$ = The nation
is united behind candidate Jones, $q$ = The nation is united behind
candidate Smith. Each of $p, q$ might be good (to avoid political
deadlock) but it can't be good for their individual goodness to
make each true since their joint truth is impossible.

This brings us to \ref{sk}\ref{skh}. The item may seem to be misclassified as
invalid if read as the inference from ``$p$ is obligatory'' to ``at least one
of $p, q$ is obligatory.'' But the appeal of the latter inference might rely on
the illicit interpretation of ``at least one of $p, q$ is obligatory'' as
``either $p$ is obligatory or $q$ is obligatory (or both), a different
assertion [see \ref{sm}\ref{smd} below].  Observe that if $\OB(p)$ entails
$\OB(p\vee q)$ then surely $\OB(\neg p)$ entails $\OB(\neg p \vee \neg q)$
hence $\PE(p\wedge q)$ entails $\PE(q)$.  But the latter assertion is open to
the same objection raised against \ref{sk}\ref{skb}. That is, it might be
permissible for a medical team to anaesthetize and make an incision without it
being permitted \textit{tout court} to make an incision; for, the incision is
forbidden in case the anaesthetic was forgotten.  Note also that the
substitution of $\neg p$ for $q$ in the conclusion of \ref{sk}\ref{skh} yields
$\OB(\TAUT)$ which might be considered objectionable [as mentioned above in
connection with \ref{ST}\ref{ST3}].

Further invalid inferences are provided by the next proposition.  Their deontic
implausibility can be easily seen from examples like those deployed above.
Some of the items in \ref{sm} have rather scant plausibility. They are included as
evidence that our logic does not ``over-generate.''

\begin{prop}\label{sm}
For every weighting function \WG:
\begin{enumerate}
\item\label{sma}\grgana{\PE p\wedge\PE q}{\PE (p\wedge q)}
\item\label{smd}\grgana{\OB (p\vee q)}{\OB p\vee \OB q}
\item\label{sme}\grgana{p\rightarrow q}{\OB p \rightarrow \OB q}
\item\label{smf}\grgana{\OB p}{\OB(p \wedge q)}
\item\label{smg}\grgana{\PE p}{\PE(p \wedge q)}
\item\label{smh}\grgana{\emptyset}{\CO(p,q)\vee \CO(q,p)}
\item\label{smi}\grgana{\OB(p\vee q)}{\OB p}
\item\label{smj}\grgana{\OB p}{p}
\item\label{smk}\grgana{\PE p}{\PE\OB p}
\item\label{sml}\grgana{(\neg q \succ\neg p) \wedge \OB p}{\OB q}
\end{enumerate}
\end{prop}

\section{Discussion}

Via the foregoing developments, we have attempted to construct deontic modality
from binary preference.  Three systems were examined. The most basic embraces
arbitrary models of the underlying preference logic and hence yields the fewest
deontic validities [e.g., those exhibited in Proposition \ref{ST}]. The $\TRI$
system narrows the class of models by identifying worlds with subsets of
propositional variables, and requiring selection to minimize the symmetric
difference between chosen and candidate worlds. New theorems emerge [see
Proposition \ref{DBProp1}].  The third system imposes an additional constraint
on selection by declaring some variables to be more stable than others when
choosing a possible world. This yields further validities [Proposition
\ref{wv}].

At the same time, many implausible deontic principles are invalid even in the
most constrained, third system [see Propositions \ref{sk}, \ref{sm}]. Indeed,
we hope to have shown that preference-based deontic logic makes tenable
distinctions between validity and invalidity, especially when the impact of an
obligation is distinguished from the impact of actually fulfilling the
obligation [as already noted in \citet{Hint}].  To illustrate the latter point
one more time, consider again \ref{sk}\ref{skz}:

\begin{quote}
$\OB(p \rightarrow q)$\ therefore\ $\OB(p) \rightarrow \OB(q)$
\end{quote}

\noindent
This inference is central to standard deontic logic \citep[\S
2.1]{sep-logic-deontic} but is invalid in our framework. Its intuitive
invalidity is revealed by the instances: 

\begin{quote}
$p$ = I promise to call this evening.

$q$ = I call this evening.
\end{quote}

\noindent
For these statements, $\OB(p \rightarrow q)$ may well be true (promise keeping).
But $\OB p$ is nonetheless insufficient to secure $\OB q$; for, $\OB(p
\rightarrow q) \wedge \OB p$ is consistent with my failing to fulfill $\OB p$
(even though it's my obligation). In this case, the obligation $\OB q$ is not
in force (or at least, not for promise reasons). Other points of divergence
between standard deontic logic and the system advanced here include the
inference from $\OB p$ to $\OB(p \vee q)$, discussed in the Introduction and in
connection with Proposition \ref{sk}\ref{skh}.

The goal, of course, is not to construct the uniquely accurate deontic logic;
there is too much variability in judgment for such a project.  Rather, we hope
to populate the space of credible systems by starting from a novel semantics,
namely, the preference semantics elaborated in \citet{PBOR}. A wide range of
deontic logics may prove helpful in analyzing obligation and related concepts,
as well as facilitating applications to Decision Science, jurisprudence, etc.

The underlying preference semantics can be enriched in several ways, each
generating a distinct deontic system. As mentioned earlier, we have here
exploited only a single utility scale whereas a plurality might be envisioned,
measuring different aspects of morality (fairness, well being of the least
favored, etc.). The question then arises how the scales should be combined; see
the discussion in \citet[\S 5.3]{PBOR}, \citet[\S 8-9]{pbbook}.  Likewise, new
constraints on models may be introduced, and the language can be enriched with
quantifiers (op.\ cit.).  In each case, a wealth of axiomatic and decidability issues
arise.

\newpage
\section*{Appendix: Remarks on demonstrations}

Proposition \ref{unrav1} is evident, and implies Proposition
\ref{ST}\ref{ST6},\ref{ST4}.  The last four items of Proposition \ref{ST} follow
from clause (\ref{defSat2a}) of Definition \ref{defSat2}.  For the harder
propositions, we provide some partial proofs.

\subsection*{Proof of Proposition \ref{DBProp1}}

We illustrate the proof of Proposition \ref{DBProp1} by doing clause
(\ref{DBProp1f}):
\begin{quote} \drgana{\OB p \wedge (\neg p \succ\neg q)}{\OB q} \end{quote}
\noindent To prove this assertion it must be shown that $\WF{\OB p \wedge (\neg
p \succ\neg q)}{\MO{M}} \subseteq \WF{\OB q}{\MO{M}}$, for all $\TRI$ models
\MO{M}.  Let $\varphi$ be $(p \succ \neg p)\wedge (\neg p \succ\neg q)$, and
let $\psi$ be $q \succ \neg q$. By Proposition \ref{unrav1}\ref{unrav1a} it
suffices to show that $\WF{\varphi}{\MO{M}}\subseteq\WF{\psi}{\MO{M}}$ for all
$\TRI$ models \MO{M}. Let $\TRI$ model $\MO{M} = \AMOD$ be given.
We partition \WORLDS\ into four cosets:
\begin{dispar}\label{app1}
\begin{enumerate}
\item\label{app1a}worlds that include both $p$ and $q$.

\item\label{app1b}worlds that include $p$ but not $q$.

\item\label{app1c}worlds that include $q$ but not $p$. and

\item\label{app1d}worlds that include neither $p$ nor $q$.
\end{enumerate}
\end{dispar}
Arbitrary worlds of these four types will be denoted by $[pq]$,
$[p\bar q]$, $[\bar p q]$, and $[\bar p \bar q]$.
We consider in turn the four kinds of worlds distinguished
in \ref{app1}. In each case it must be shown that if the
world belongs to $\WF{\varphi}{\MO{M}}$ then it belongs
to $\WF{\psi}{\MO{M}}$.

To begin, choose a world $[pq]$ of type \ref{app1}\ref{app1a}.
Because $\SG$ is $\TRI$-based, we have:

\begin{dispar}\label{apa1}
\begin{enumerate}
\item\label{apa1a} $\SG([pq],\WF{p}{\MO{M}}) = [pq]$
\item\label{apa1b} $\SG([pq],\WF{q}{\MO{M}}) = [pq]$
\item\label{apa1c} $\SG([pq],\WF{\neg p}{\MO{M}}) = [\bar pq]$
\item\label{apa1d} $\SG([pq],\WF{\neg q}{\MO{M}}) = [p\bar q]$
\end{enumerate}
\end{dispar}
where for all variables $v$ not equal to $p$ or $q$, $v$ belongs to
one of $[pq]$, $[p\bar q]$, $[\bar p q]$ iff $v$ belongs to all three
of them. Suppose that $[pq] \in \WF{\varphi}{\MO{M}}$. Then:

\begin{dispar}\label{apa2}
\begin{enumerate}
\item\label{apa2a}
$\UG(\SG([pq], \WF{\neg p}{\MO{M}})) > \UG(\SG([pq],\WF{\neg q}{\MO{M}}))$
\item\label{apa2b}
$\UG(\SG([pq],\WF{p}{\MO{M}})) > \UG(\SG([pq],\WF{\neg p}{\MO{M}}))$
\end{enumerate}
\end{dispar}
So by \ref{apa1} and \ref{apa2}:

\begin{dispar}\label{apa3}
\begin{enumerate}
\item\label{apa3a}
$\UG([\bar p q]) > \UG([p\bar q])$
\item\label{apa3b}
$\UG([pq])>\UG([\bar p q])$
\end{enumerate}
\end{dispar}
Directly from \ref{apa3}:
\begin{quote}
$\UG([pq]) > \UG([p\bar q])$,
\end{quote}
hence by \ref{apa1}:
\begin{dispar}\label{apa4}
$\UG(\SG([pq],\WF{q}{\MO{M}})) > \UG(\SG([pq],\WF{\neg q}{\MO{M}}))$
\end{dispar}
And \ref{apa4} implies that $[pq]\in\WF{\psi}{\MO{M}}$.


Next, choose a world $[p\bar q]$ of type \ref{app1}\ref{app1b}.
Because $\SG$ is $\TRI$-based, we have:

\begin{dispar}\label{apa5}
\begin{enumerate}
\item\label{apa5a} $\SG([p\bar q],\WF{p}{\MO{M}}) = [p\bar q]$
\item\label{apa5b} $\SG([p\bar q],\WF{q}{\MO{M}}) = [pq]$
\item\label{apa5c} $\SG([p\bar q],\WF{\neg p}{\MO{M}}) = [\bar p\bar q]$
\item\label{apa5d} $\SG([p\bar q],\WF{\neg q}{\MO{M}}) = [p\bar q]$
\end{enumerate}
\end{dispar}
where for all variables $v$ not equal to $p$ or $q$, $v$ belongs to
one of $[pq]$, $[p\bar q]$, $[\bar p \bar q]$ iff $v$ belongs to all three
of them. Suppose that $[p\bar q] \in \WF{\varphi}{\MO{M}}$. Then:

\begin{dispar}\label{apa7}
\begin{enumerate}
\item\label{apa7a}
$\UG(\SG([p\bar q], \WF{\neg p}{\MO{M}})) > \UG(\SG([p\bar q],\WF{\neg q}{\MO{M}}))$
\item\label{apa7b}
$\UG(\SG([p\bar q],\WF{p}{\MO{M}})) > \UG(\SG([p\bar q],\WF{\neg p}{\MO{M}}))$
\end{enumerate}
\end{dispar}
So by \ref{apa5} and \ref{apa7}:

\begin{dispar}\label{apa8}
\begin{enumerate}
\item\label{apa8a}
$\UG([\bar p\bar q]) > \UG([p\bar q])$
\item\label{apa8b}
$\UG([p\bar q]) > \UG([\bar p \bar q])$
\end{enumerate}
\end{dispar}
But \ref{apa8} is impossible, contradicting
$[p\bar q] \in \WF{\varphi}{\MO{M}}$. Thus,
$[p\bar q]$ is not a counter example to
$\WF{\varphi}{\MO{M}}\subseteq\WF{\psi}{\MO{M}}$.


Next, choose a world $[\bar pq]$ of type \ref{app1}\ref{app1c}.
Because $\SG$ is $\TRI$-based, we have:

\begin{dispar}\label{apa10}
\begin{enumerate}
\item\label{apa10a} $\SG([\bar pq],\WF{p}{\MO{M}}) = [pq]$
\item\label{apa10b} $\SG([\bar pq],\WF{q}{\MO{M}}) = [\bar pq]$
\item\label{apa10c} $\SG([\bar pq],\WF{\neg p}{\MO{M}}) = [\bar p q]$
\item\label{apa10d} $\SG([\bar pq],\WF{\neg q}{\MO{M}}) = [\bar p\bar q]$
\end{enumerate}
\end{dispar}
where for all variables $v$ not equal to $p$ or $q$, $v$ belongs to
one of $[pq]$, $[\bar p q]$, $[\bar p \bar q]$ iff $v$ belongs to all three
of them. Suppose that $[\bar pq] \in \WF{\varphi}{\MO{M}}$. Then:

\begin{dispar}\label{apa12}
\begin{enumerate}
\item\label{apa12a}
$\UG(\SG([\bar pq], \WF{\neg p}{\MO{M}})) > \UG(\SG([\bar pq],\WF{\neg q}{\MO{M}}))$
\item\label{apa12b}
$\UG(\SG([\bar pq],\WF{p}{\MO{M}})) > \UG(\SG([\bar pq],\WF{\neg p}{\MO{M}}))$
\end{enumerate}
\end{dispar}
So by \ref{apa10} and \ref{apa12}:

\begin{dispar}\label{apa13}
\begin{enumerate}
\item\label{apa13a}
$\UG([\bar p q]) > \UG([\bar p\bar q])$
\item\label{apa13b}
$\UG([pq])>\UG([\bar p q])$
\end{enumerate}
\end{dispar}
Directly from \ref{apa13}\ref{apa13a} and \ref{apa10}\ref{apa10b},\ref{apa10d}:
\begin{dispar}\label{apa14}
$\UG(\SG([\bar p q],\WF{q}{\MO{M}})) > \UG(\SG([\bar p q],\WF{\neg q}{\MO{M}}))$
\end{dispar}
And \ref{apa14} implies that $[\bar p q]\in\WF{\psi}{\MO{M}}$.


Finally, choose a world $[\bar p\bar q]$ of type \ref{app1}\ref{app1d}.
Because $\SG$ is $\TRI$-based, we have:

\begin{dispar}\label{apa9}
\begin{enumerate}
\item\label{apa9a} $\SG([\bar p\bar q],\WF{p}{\MO{M}}) = [p\bar q]$
\item\label{apa9b} $\SG([\bar p\bar q],\WF{q}{\MO{M}}) = [\bar pq]$
\item\label{apa9c} $\SG([\bar p\bar q],\WF{\neg p}{\MO{M}}) = [\bar p\bar q]$
\item\label{apa9d} $\SG([\bar p\bar q],\WF{\neg q}{\MO{M}}) = [\bar p\bar q]$
\end{enumerate}
\end{dispar}
where for all variables $v$ not equal to $p$ or $q$, $v$ belongs to
one of $[p\bar q]$, $[\bar p q]$, $[\bar p \bar q]$ iff $v$ belongs to all three
of them. Suppose that $[\bar p\bar q] \in \WF{\varphi}{\MO{M}}$. Then:

\begin{dispar}\label{apa91}
\begin{enumerate}
\item\label{apa91a}
$\UG(\SG([\bar p\bar q], \WF{\neg p}{\MO{M}})) > \UG(\SG([\bar p\bar q],\WF{\neg q}{\MO{M}}))$
\item\label{apa91b}
$\UG(\SG([\bar p\bar q], \WF{p}{\MO{M}})) > \UG(\SG([\bar p\bar q],\WF{\neg p}{\MO{M}}))$
\end{enumerate}
\end{dispar}
So by \ref{apa9}\ref{apa9c},\ref{apa9d} and \ref{apa91}\ref{apa91a}:
\begin{quote}
$\UG([\bar p\bar q]) > \UG([\bar p\bar q])$.
\end{quote}
But this is impossible, contradicting
$[\bar p\bar q] \in \WF{\varphi}{\MO{M}}$. Thus,
$[\bar p\bar q]$ is not a counter example to
$\WF{\varphi}{\MO{M}}\subseteq\WF{\psi}{\MO{M}}$. \QED

\subsection*{Proof of Proposition \ref{wv}}

We illustrate the proof of Proposition \ref{wv} by doing clause
(\ref{wvh}):

\begin{quote}
\ergana{\OB(p\vee q)\wedge \neg q}{\OB p}
\end{quote}

\noindent
where $\WGC$ is the class of weighting function that assign greater value to
$q$ than to either $p$ or $r$.  To prove the assertion it must be shown that
$\WF{\OB(p\vee q)\wedge \neg q }{\MO{M}} \subseteq \WF{\OB p}{\MO{M}}$, for all
$\TRI$ models with $\WG$-based selection function drawn from $\WGC$.  Let
$\varphi$ be $(((p\vee q) \succ (\neg p\wedge \neg q))\wedge \neg q)$, and let
$\psi$ be $p \succ \neg p$. By Proposition \ref{unrav1}\ref{unrav1a} it
suffices to show that $\WF{\varphi}{\MO{M}}\subseteq\WF{\psi}{\MO{M}}$ for all
$\TRI$ models with $\WG$-based selection function drawn from $\WGC$. Let
$\MO{M}$ be such a model.  We partition \WORLDS\ into the four cosets indicated
in \ref{app1}, and denote their members using the same notation as before.
Observe that for all worlds $w$ of types $[pq]$ or $[\bar p q]$,
$w\not\in\WF{\varphi}{\MO{M}}$. It thus suffices to show that if $[p\bar
q]\in\WF{\varphi}{\MO{M}}$ then $[p\bar q]\in\WF{\psi}{\MO{M}}$, and that if
$[\bar p\bar q]\in\WF{\varphi}{\MO{M}}$ then $[\bar p\bar q]
\in\WF{\psi}{\MO{M}}$. First consider $[p\bar q]$.
Because $\SG$ is \WGC-based (and hence $\TRI$), we have:

\begin{dispar}\label{apb1}
\begin{enumerate}
\item\label{apb1a} $\SG([p\bar q],\WF{\neg p \wedge \neg q}{\MO{M}}) = [\bar p\bar q]$
\item\label{apb1b} $\SG([p\bar q],\WF{p}{\MO{M}}) = [p\bar q]$
\item\label{apb1c} $\SG([p\bar q],\WF{\neg p}{\MO{M}}) = [\bar p\bar q]$
\item\label{apb1d} $\SG([p\bar q],\WF{p \vee q}{\MO{M}}) = [p\bar q]$
\end{enumerate}
\end{dispar}
where for all variables $v$ not equal to $p$ or $q$, $v$ belongs to
one of $[\bar p\bar q]$, $[p\bar q]$ iff $v$ belongs to both
of them. Suppose that $[p\bar q] \in \WF{\varphi}{\MO{M}}$. Then:

\begin{dispar}\label{apb2}
$\UG(\SG([p\bar q], \WF{p\vee q}{\MO{M}})) > \UG(\SG([p\bar q],\WF{\neg p \wedge \neg q}{\MO{M}}))$
\end{dispar}
So by \ref{apb1} and \ref{apb2}:
\begin{dispar}\label{apb3}
$\UG([p\bar q]) > \UG([\bar p\bar q])$
\end{dispar}
hence by \ref{apb1}:
\begin{dispar}\label{apb4}
$\UG(\SG([p\bar q],\WF{p}{\MO{M}})) > \UG(\SG([p\bar q],\WF{\neg p}{\MO{M}}))$
\end{dispar}
And \ref{apb4} implies that $[p\bar q]\in\WF{\psi}{\MO{M}}$.

Now consider $[\bar p\bar q]$.
Because $\SG$ is \WGC-based, we have:

\begin{dispar}\label{apc1}
\begin{enumerate}
\item\label{apc1a} $\SG([\bar p\bar q],\WF{\neg p \wedge \neg q}{\MO{M}}) = [\bar p\bar q]$
\item\label{apc1b} $\SG([\bar p\bar q],\WF{p}{\MO{M}}) = [p\bar q]$
\item\label{apc1c} $\SG([\bar p\bar q],\WF{\neg p}{\MO{M}}) = [\bar p\bar q]$
\item\label{apc1d} $\SG([\bar p\bar q],\WF{p \vee q}{\MO{M}}) = [p\bar q]$
\end{enumerate}
\end{dispar}
where for all variables $v$ not equal to $p$ or $q$, $v$ belongs to
one of $[\bar p\bar q]$, $[p\bar q]$ iff $v$ belongs to both
of them. Note that \ref{apc1}\ref{apc1d} relies on our choice of weighting function.
Since the function gives greater weight to $q$ than to $p$, the polarity
of $q$ is preserved rather than that of $p$ when a choice between the two must be made.

Suppose that $[\bar p\bar q] \in \WF{\varphi}{\MO{M}}$. Then:
\begin{dispar}\label{apc2}
$\UG(\SG([\bar p\bar q], \WF{p\vee q}{\MO{M}})) > \UG(\SG([\bar p\bar q],\WF{\neg p \wedge \neg q}{\MO{M}}))$
\end{dispar}
So by \ref{apc1} and \ref{apc2}:
\begin{dispar}\label{apc3}
$\UG([p\bar q]) > \UG([\bar p\bar q])$
\end{dispar}
hence by \ref{apc1}:
\begin{dispar}\label{apc4}
$\UG(\SG([\bar p\bar q],\WF{p}{\MO{M}})) > \UG(\SG([\bar p\bar q],\WF{\neg p}{\MO{M}}))$
\end{dispar}
And \ref{apc4} implies that $[\bar p\bar q]\in\WF{\psi}{\MO{M}}$. \QED

\subsection*{Proof of Propositions \ref{sk} and \ref{sm}}

We do \ref{sk}\ref{skz}:
\begin{quote}
For every weighting function \WG:
\grgana{\OB(p\rightarrow q)}{\OB(p) \rightarrow\OB(q)}
\end{quote}

\noindent
Let $\varphi = (p \rightarrow q) \succ (p \wedge \neg q)$ and $\psi = (p \succ
\neg p) \rightarrow (q \succ \neg q)$.  By Proposition
\ref{unrav1}\ref{unrav1a} it suffices to show that for every weighting function
$\WG$, there is $\TRI$ model $\MO{M} = \AMOD$ with $\WG$-based selection
function such that for some $w\in\WORLDS$: $w\in\WF{\varphi}{\MO{M}}$ but
$w\not\in\WF{\psi}{\MO{M}}$. We partition \WORLDS\ into the four cosets
indicated in \ref{app1}, and denote their members using the same notation as
before.  Pick a world of form $[\bar p q]$, Let $\MO{M} = \AMOD$ be such that:

\begin{dispar}\label{apd1}
\begin{enumerate}
\item\label{apd1a} $\SG([\bar p q],\WF{p \rightarrow q}{\MO{M}}) = [\bar p q]$
\item\label{apd1b} $\SG([\bar p q],\WF{p \wedge \neg q}{\MO{M}}) = [ p\bar q]$
\item\label{apd1c} $\SG([\bar p q],\WF{p}{\MO{M}}) = [p q]$
\item\label{apd1d} $\SG([\bar p q],\WF{\neg p}{\MO{M}}) = [\bar p q]$
\item\label{apd1e} $\SG([\bar p q],\WF{q}{\MO{M}}) = [\bar p q]$
\item\label{apd1f} $\SG([\bar p q],\WF{\neg q}{\MO{M}}) = [\bar p\bar q]$
\end{enumerate}
\end{dispar}

\noindent
where for all variables $v$ not equal to $p$ or $q$, $v$ belongs to one of
$[\bar p q]$, $[p\bar q]$, $[pq]$, $[\bar p \bar q]$ iff $v$ belongs to all
four of them.  It is clear from \ref{apd1} that $\SG$ may be chosen to be
$\WG$-based for any weighting function $\WG$ inasmuch as all of the selections
are dictated exclusively by minimizing symmetric difference of variables
in the sense of subset. (See Section \ref{SS:prox} above.)

Let utility function $\UG$ be such that:
\begin{dispar}\label{apd2}
\begin{enumerate}
\item\label{apd2a} $\UG(\SG([\bar p q],\WF{p \rightarrow q}{\MO{M}})) = \UG([\bar p q])$ = $2$
\item\label{apd2b} $\UG(\SG([\bar p q],\WF{p \wedge \neg q}{\MO{M}})) = \UG([ p\bar q])$ = $1$
\item\label{apd2c} $\UG(\SG([\bar p q],\WF{p}{\MO{M}})) = \UG([p q])$ = $3$
\item\label{apd2d} $\UG(\SG([\bar p q],\WF{\neg p}{\MO{M})}) = \UG([\bar p q])$ = $2$
\item\label{apd2e} $\UG(\SG([\bar p q],\WF{q}{\MO{M}})) = \UG([\bar p q])$ = $2$
\item\label{apd2f} $\UG(\SG([\bar p q],\WF{\neg q}{\MO{M})}) = \UG([\bar p\bar q])$ = $4$
\end{enumerate}
\end{dispar}

\noindent
Then 
$\UG(\SG([\bar p q],\WF{p \rightarrow q}{\MO{M}})) = 2 > 1 = \UG(\SG([\bar p q],\WF{p \wedge \neg q}{\MO{M}}))$
so $[\bar p q] \in\WF{\varphi}{\MO{M}}$. On the other hand,
$\UG(\SG([\bar p q],\WF{p}{\MO{M}})) = 3 > 2 = \UG(\SG([\bar p q],\WF{\neg p}{\MO{M})})$ but
$\UG(\SG([\bar p q],\WF{q}{\MO{M}})) = 2 < 4 = \UG(\SG([\bar p q],\WF{\neg q}{\MO{M})})$.
Hence, $[\bar p q]\not\in\WF{\psi}{\MO{M}}$. \QED

\clearpage

\renewcommand{\baselinestretch}{1.0}

\end{document}